\documentclass[twocolumn,pre,floatfix,showpacs]{revtex4}

\usepackage{epsfig}
\usepackage{amsmath}
\usepackage{times}

\usepackage{epsfig}

\begin{document}

\title{How would you integrate the equations of motion in 
       dissipative particle dynamics simulations?}
\date{\today}

\author{P. Nikunen and M. Karttunen}
\affiliation{Biophysics and Statistical Mechanics Group, 
Laboratory of Computational Engineering, Helsinki University of Technology, 
P.O. Box 9203, FIN--02015 HUT, Finland} 
\author{I. Vattulainen}
\affiliation{Laboratory of Physics and Helsinki Institute of Physics, 
Helsinki University of Technology, P.O. Box 1100, FIN--02015 HUT, Finland}

\begin{abstract} 

In this work we assess the quality and performance of several 
novel dissipative particle dynamics integration schemes that 
have not previously been tested independently. Based on 
a thorough comparison we identify the respective methods of 
Lowe and Shardlow as particularly promising candidates for 
future studies of large-scale properties of soft matter systems. 

\end{abstract}



\maketitle

\section{Introduction}

The mesoscopic phenomena of so-called ``soft matter'' physics 
\cite{Dao99,Cat00,deGennes96}, embracing a diverse range of 
systems including liquid crystals, colloids, and biomembranes,
generally involve some form of coupling between different 
characteristic time- and length-scales. Computational modeling 
of such multi-scale effects requires new methodology applicable 
beyond the realm of traditional techniques such as {\it ab initio} 
and classical molecular dynamics \cite{Fre02,Ber98} (the methods 
of choice in the microscopic regime), and phase field modeling 
\cite{Eld01} or the lattice-Boltzmann method \cite{Rot97} 
(usually concerned with the macroscopic regime).

Dissipative particle dynamics (DPD) \cite{Hoo92,Esp95,War98,Gro97} 
is a particularly appealing technique in this regard. 
The ``particles'' of DPD correspond to coarse-grained entities,
representing a collection of molecules or molecular groups rather 
than individual atoms. Coarse-graining leads to soft pair 
potentials allowing the particles to overlap 
(Forrest and Suter \cite{For95}).

Although coarse-graining might also be considered implicit in 
Brownian and Langevin dynamics simulation, DPD offers the explicit 
advantage of a proper description of hydrodynamic modes significant 
in the physical approach towards a system's equilibrium. This is 
achieved in DPD by implementing a thermostat in terms of pairwise 
random and dissipative forces such that the total momentum of the 
system is conserved.
Due to these reasons, DPD has been used in studies covering
a wide range of aspects in soft matter systems, including the 
structure of lipid bilayers \cite{Ven99,Gro01}, self-assembly 
\cite{Yam02}, and the formation of polymer-surfactant complexes 
\cite{Gro00}.

In practice, the pairwise coupling of particles through 
random and dissipative forces makes the integration of the 
equations of motion a nontrivial task. The main difficulty 
arises from the dissipative force, which depends explicitly on 
the relative velocities of the particles, while the velocities 
in turn depend on the dissipative forces. An accurate 
description of the dynamics requires a self-consistent 
solution. 

The considerable computational load associated with 
this task has motivated the development of schemes 
\cite{Gro97,Nov98,Gib99,Pag98,Bes00,Vat02,Ott01,Sha01} 
which account for the velocity dependence of dissipative forces 
in some approximate manner, allowing the integration to be 
carried out to a sufficient degree of computational efficiency.
The search for a satisfactory such integration scheme is ongoing,
since many of the recent proposal have been found to exhibit 
non-physical behavior, such as  systematic drift in temperature, 
and artificial structures in the radial distribution 
function \cite{Pag98,Bes00,Vat02}.

In order to overcome these problems, a number of new integration 
schemes for DPD simulations have been developed in the past few 
years. Self-consistent determination schemes exist on the one 
hand \cite{Pag98,Bes00,Vat02}, but these are rather elaborate.

Alternative proposals include (i) a parameterization of the 
integrator based on the specific application being modeled by 
den Otter and Clarke \cite{Ott01}, (ii) operator splitting by 
Shardlow \cite{Sha01}, and (iii) an elegant Monte Carlo-based 
approach due to Lowe \cite{Low99} which completely avoids the 
problems arising from random and dissipative forces as it does 
not use random or dissipative forces at all.

In this article, we apply these schemes respectively to specific
model systems, with the objective of assessing their relative 
performance. To the authors' knowledge, the latter three have 
yet to be tested and compared independently. 
The self-consistent approach has been tested, although not 
extensively, and the previously tested so-called DPD-VV 
(DPD version of the ``velocity-Verlet'' scheme) will be 
used here as a benchmark. 
By monitoring a number of physical observables including
temperature, radial distribution function, radius of 
gyration for polymers, and tracer diffusion, we 
find that the methods by Lowe \cite{Low99} and 
Shardlow \cite{Sha01} give the best overall performance
and are superior also to the integrators tested 
in previous studies \cite{Bes00,Vat02}. As will be 
discussed in the last section, a direct comparison 
between these two is not straightforward, since 
they are based on essentially different conceptual 
views of dissipative particle dynamics.

The paper is organized as follows. First, we briefly summarize the
dissipative particle dynamics method and introduce the three model systems
used here. In Sect.~\ref{sec:integrators} we describe the integration
algorithms and the update schemes in detail, and in  Sect.~\ref{sec:per}
we present the results from the tests. Finally, in Sect.~\ref{sec:sum}
the findings and their relevance are discussed.

\section{Methods and Models}
\label{sec:met}

Below we give a short summary of the dissipative particle dynamics 
method and describe the three model systems used in this work. 
For more thorough accounts on DPD, see e.g. Refs.~\cite{War98,Gro97}.

\subsection{Dissipative Particle Dynamics}

Dissipative particle dynamics describes a system in terms of 
$N$ particles having masses $m_i$, positions $\vec{r}_i$, 
and velocities $\vec{v}_i$. Interactions are composed of 
pairwise conservative, dissipative, and random forces exerted 
on particle $i$ by particle $j$, respectively, and are given by 
  \begin{equation}
  \begin{array}{lcl} 
  \vec{F}_{ij}^C & = & F_{ij}^{(c)}(r_{ij}) \, \vec{e}_{ij}, \\
  \vec{F}_{ij}^D & = & - \gamma \, \omega^D(r_{ij}) \, 
                          (\vec{v}_{ij} \cdot \vec{e}_{ij})
                          \, \vec{e}_{ij}, \\
  \vec{F}_{ij}^R & = & \sigma \, \omega^R(r_{ij}) \, \xi_{ij} \, \vec{e}_{ij} ,
  \end{array} 
  \label{eq:frij}
  \end{equation}
where $\vec{r}_{ij} \equiv \vec{r}_i - \vec{r}_j$, 
$r_{ij} \equiv | \vec{r}_{ij} |$, 
$\vec{e}_{ij} \equiv \vec{r}_{ij}/r_{ij}$, and 
$\vec{v}_{ij} \equiv \vec{v}_i - \vec{v}_j$. The variables 
$\gamma$ and $\sigma$ are the strengths of the dissipative and 
random forces, respectively. The $\xi_{ij}$ are symmetric Gaussian
random variables with zero mean and unit variance, and are independent 
for different pairs of particles and different times. The condition 
$\xi_{ij} = \xi_{ji}$ is employed to ensure momentum conservation.

The pairwise conservative force $F_{ij}^{(c)}$ 
is not specified by the DPD formulation and it can be chosen 
to include any forces that are appropriate for a given system, 
such as van der Waals and electrostatic interactions.
In addition, it is important to notice that it is completely 
independent of the random and dissipative forces. 
Since one of the main motivations for using DPD is to be able
to simulate systems at coarse-grained, or mesoscopic, scales,
the conservative force is often chosen to be soft repulsive. 
Here, we use the ``classical'' DPD choice, i.e., soft repulsive 
conservative forces in the cases of model A and model B, and 
hard Lennard-Jones interactions combined with harmonic spring 
forces in the model polymer system.

In contrast to the conservative force, the random and dissipative 
forces are not independent, but are 
coupled through a fluctuation-dissipation relation. This 
coupling is due to the requirement that in thermodynamic 
equilibrium the system must  have canonical distribution. 
The necessary conditions were first derived by Espa{\~n}ol 
and Warren in 1995 using a Fokker-Planck equation \cite{Esp95}.

The requirement of canonical distribution sets 
two conditions linking the random and dissipative forces
in Eq.~(\ref{eq:frij}). The first one couples the weight 
functions through $\omega^D(r_{ij}) = [\omega^R(r_{ij})]^2$, 
and the second one the strengths of the random and dissipative forces 
via $\sigma^2 = 2 \gamma k_B T^*$. The latter condition  fixes 
the temperature of the system $T^*$  ($k_B$ being the Boltzmann constant)
and relates it to the two DPD parameters $\gamma$ and $\sigma$.

Like classical molecular dynamics (MD) simulations, DPD allows 
studies of dynamical properties since the time 
evolution of particles can be described by the 
Newton's equations of motion
  \begin{equation}
  \begin{array}{lcl} 
  d \vec{r}_i & = & \vec{v}_i \, dt, \\
  d \vec{v}_i & = & \frac{1}{m_i} ( \vec{F}_i^C dt + \vec{F}_i^D dt +
  \vec{F}_i^R \sqrt{dt} ) . 
  \end{array} 
  \label{eq:newton2}
  \end{equation}
Here $\vec{F}_i^C = \sum_{i\neq j}\vec{F}_{ij}^C$ is the total 
conservative force acting on particle $i$, and $\vec{F}_i^D$ and 
$\vec{F}_i^R$ are defined correspondingly. It is important to 
notice that the velocity 
increment due to the random force in Eq.~(\ref{eq:newton2})
has the factor $\sqrt{dt}$ instead of $dt$. 
It can be justified by a Wiener process as in stochastic 
differential equations. Here, it suffices to notice that physically
the Wiener process models intrinsic (thermal) noise in the system and
provides the simplest approach to modeling  Brownian motion using
stochastic processes (see Refs.~\cite{Esp95,Gro97} for a detailed 
discussion). The above continuous-time version of DPD satisfies 
detailed balance and describes the canonical NVT ensemble. In 
practice, however, the time increments in Eq.~(\ref{eq:newton2}) 
are finite and the equations of motion must be solved by 
some integration procedure. We will return to this issue in 
Sect.~\ref{sec:integrators}.

\subsection{An alternative approach to DPD}

Dissipative particle dynamics described above can be thought 
of as a momentum conserving thermostat that allows one to study 
a system within the NVT ensemble with full hydrodynamics. The 
key features are therefore momentum and temperature conservation. 
As discussed above, momentum conservation arises naturally 
from pairwise forces. Temperature conservation, in turn, 
arises from the random and dissipative forces that are 
chosen to satisfy the fluctuation-dissipation theorem.

An alternative approach was formulated by Lowe in 1999 
\cite{Low99}. It does not use dissipative or random forces at 
all, yet provides the same conservation laws and is similar in 
spirit to DPD as it is aimed for studies of coarse-grained models 
in terms of soft interactions. In Lowe's method, one first 
integrates Newton's equations of motion with a time step 
$\Delta t$, and then thermalizes the system using the Andersen 
thermostat \cite{And80} for pairs of particles. We will discuss 
this method in detail in Sect.~\ref{sect:lowes_integrator}.

Lowe's approach is appealing for a number of reasons. First 
of all, since there are no dissipative forces we can assume that 
this  method does not suffer from the same drawback as DPD: 
While DPD requires a self-consistent solution of the equations 
of motion, Lowe's approach is easier to use and 
most likely performs well even with integration schemes that are 
commonly used in classical MD simulations. 
Secondly, the rate of how often the particle velocities are 
thermalized may be varied over a wide range, which implies that 
the dynamical properties of the system can be tuned in a controlled 
fashion. Lowe has demonstrated this idea 
by showing how 
some dimensionless variables (such as the Schmidt number $Sc$) 
can be tuned to match values found in actual fluids \cite{Low99}.

\subsection{Model systems}

In this study, we evaluate the performance of a number of 
novel integration schemes that have been recently suggested 
for large-scale DPD simulations
(see Sect.~\ref{sec:integrators}). We test these integrators 
using three different model systems. The first two are based 
on a 3D model fluid system with a fixed number of identical 
particles. The first of them is aimed to clarify the 
performance of integration schemes in weakly interacting 
systems dominated by the random and dissipative forces, while 
the second model is relevant for systems in which the 
conservative interactions are of major importance. Finally, to 
gain insight into problems associated with hybrid models in 
which both soft and hard interactions are included, we consider 
a model of an individual polymer chain in a hydrodynamic solvent.

\subsubsection{Model A}

Model A describes  the case characterized by the absence of 
conservative forces ($F_{ij}^{(c)} = 0$). This choice corresponds 
to an ideal gas and it is customarily called ``ideal DPD fluid''.
The reason for using this model is that it provides us with some 
exact theoretical results that can be compared to results from 
model simulations. Here, 
the dynamics of the system arises only from thermal noise and 
dissipative coupling between pairs of particles. In DPD 
simulations, the random force strength is chosen to be $\sigma = 3$ 
in units of $k_BT^{\ast}$, and the strength of the dissipative 
force $\gamma$ is then determined by the fluctuation-dissipation 
relation $\sigma^2 = 2\gamma k_BT^{\ast}$.

The random and dissipative forces are chosen to be soft-repulsive, 
  \begin{equation}
  \label{eq:omega_rij}
  \omega(r_{ij}) = \left\{
  \begin{array}{ll}
    1 - r_{ij}/r_c & \mbox{for $r_{ij}<r_c$}; \\
    0 & \mbox{for $r_{ij}>r_c$},
  \end{array}
  \right.
  \end{equation}
with a cut-off distance $r_c$ \cite{Gro97} and 
$\omega^{R}(r_{ij}) = \omega(r_{ij})$. This is the most common 
choice in DPD simulations and it has been used 
in recent investigations of integration schemes 
\cite{Bes00,Vat02}, thus allowing for a comparison of the present 
results with those of previous works. Although 
Eq.~(\ref{eq:omega_rij}) has been used in virtually all
published studies using DPD, it should be noted that 
the fluctuation-dissipation theorem does not specify the 
functional form of the weight function. The simple
form of Eq.~(\ref{eq:omega_rij}) simply provides a convenient choice.

In our simulations, a 3D simulation box of size $10\times 10\times 10$ 
with periodic boundary conditions is used. The length scale is defined 
by setting $r_c = 1$, and a particle number density $\rho = 4$. The energy 
scale is defined by setting the desired thermal energy to unity via 
$k_BT^{\ast} = 1$. All particles are identical, and thus 
$m_i = m$ for all $i$.

\subsubsection{Model B} 

Model B is a simple interacting DPD fluid. Its main difference 
to model A is the presence of a conservative force, 
which we choose to be of the form 
$F_{ij}^{(c)}(r_{ij}) = \mathcal{A} \, \omega(r_{ij})$.
The amplitude of the force was chosen to be
$\mathcal{A} = 25$. This functional form 
for the conservative force is by far the most common choice
in DPD simulations.

\subsubsection{Model polymer system}

The last model system considered in this work describes an 
individual polymer chain in an explicit hydrodynamic solvent. 
Our interest in a system of this kind 
originates from the fact that various soft matter systems 
such as liquid crystals and lipid bilayers are composed 
of particles which are essentially chain-like molecules. 
DPD serves well for studies of these systems 
due to the hydrodynamic nature of the solvent which plays 
an important role in various soft matter systems. However, 
while it is often desirable to describe chain-like molecules 
on a molecular level by hard interactions, complemented with 
bending and torsional potentials to account for the most 
relevant microscopic degrees of freedom, the solvent can 
often be described on a simpler level in terms of 
soft pair potentials. Thus, one possibility for efficiently 
modeling polymeric systems is a hybrid approach of chain-like 
molecules in a coarse-grained solvent.

The gain of using a hybrid approach in complex macromolecular 
systems is evident. It allows one to reduce the computational 
burden of dealing with an explicit solvent, while the molecular 
description of macromolecules is still accounted for in 
detail. However, the practical 
implications of including both hard and soft interactions in 
a model are not well understood. In a previous study for an 
ensemble of spherical particles described by hard conservative 
and soft dissipative forces, we found certain features which 
differentiated integration schemes from each other \cite{Vat02}. 
However, a full study of the performance of integration schemes 
within a true hybrid approach of a macromolecular system has 
been lacking up till now.

This model system aims to quantify 
the effects of integration schemes under conditions that combine 
both soft and hard interactions for a model polymer system. Here, 
the idea is to optimize the efficiency of the model by using 
a minimal approach. We thus describe the solvent as an ensemble 
of identical particles which interact via soft pairwise forces and 
satisfy momentum conservation, while the polymer chain is described 
on a more microscopic level in terms of (hard) Lennard-Jones 
interactions and harmonic springs.

The linear polymer chain is described as a chain of $M$ 
monomers connected by harmonic bonds whose potential 
follows the form 
\begin{equation} 
U_{\rm harm} = \frac{k}{2} | \vec{r}_i - \vec{r}_{i-1} |^2, 
\,\,\,\,\,\,\, i = 2, 3, \ldots , M ,  
\end{equation} 
with a spring constant $k = 7$. 
Within the chain, the conservative monomer-monomer interactions 
are given by the truncated and shifted Lennard-Jones potential
\begin{equation}
\label{eq:lennardjones}
U_{\rm LJ} (r_{ij}) = \left\{
\begin{array}{ll}
4\,\epsilon \left[ \left(\frac{\ell}{r_{ij}}\right)^{12}
                 - \left(\frac{\ell}{r_{ij}}\right)^{6} + \frac{1}{4}
                \right]
   & ,\, r_{ij} \leq r_c \\
0  & ,\, r_{ij} > r_c
\end{array}
\right .
\end{equation}
such that the potential is purely repulsive and decays 
smoothly to zero at $r_c$. We choose $\ell = 2^{-1/6}$ and 
$\epsilon = k_B T^{\ast}$, and therefore $ r_c = \ell \, 2^{1/6} = 1$.
The pairwise conservative force acting on a monomer due to 
other monomers in a chain therefore follows directly from
$\vec{F}^C = - \nabla ( U_{\rm harm} + U_{\rm LJ} ) $. The 
dissipative and random forces acting on the monomers are 
chosen to follow Eqs.~(\ref{eq:frij}) and (\ref{eq:omega_rij}) 
with $\sigma = 3$.

The monomer-solvent and solvent-solvent interactions are 
described as in model A with $\sigma = 3$ (and 
$\mathcal{A} = 0$), i.e., the random and dissipative
parts are used as a momentum conserving thermostat.

The justification for 
this choice of interactions lies in 
a wish to clarify the size of artifacts due to integration 
schemes in a case, where the polymer chain is described on 
a molecular level in terms of hard interactions, while the 
solvent is coarse grained as much as possible and is thus 
described by an ideal gas. The coupling between the solvent 
and the polymer chain comes from the dissipative forces 
which give rise to hydrodynamic modes. This eventually 
results in a minimal model of a polymer chain with full 
hydrodynamics under good solvent conditions.  This was
confirmed by studying the scaling behavior 
of the radius of gyration.

We consider polymers of size $M = 20$ and use a 3D box 
of size $10\times 10\times 10$ with periodic boundary conditions, 
where the length scale is defined by $r_c = 1$. The particle number 
density is chosen to be $\rho = 4$ while the energy scale is defined 
by setting the desired thermal energy to unity via $k_BT^{\ast} = 1$. 
All particles are identical, and thus $m_i = m$ for both solvent 
and monomer particles.

\subsubsection{Choice of random numbers} 

In the present work, uniformly 
distributed random numbers $u \in U(0,1)$ are used such that 
$\xi_{ij} = \sqrt{3}(2u - 1)$. This approach is highly efficient 
and yields results that are indistinguishable from those 
generated by Gaussian random numbers \cite{Gro97}. However, 
in the case of Lowe's approach, the $\xi_{ij}^{(g)}$ used 
are true Gaussian random numbers.

\section{Integrators}
\label{sec:integrators}

The integration schemes tested in this work have been 
chosen from the most recent ones that have been suggested 
in the literature but not tested and compared to other methods. 
They complement each other in the sense 
that the velocity dependence of the dissipative forces is 
accounted for in all cases, but the approaches 
differ substantially. 
We feel that all of the integrators considered here are 
promising candidates for large-scale simulations of soft matter 
systems. However, due to the lack of comparative studies in which 
all of these schemes would have been tested on equal footing, 
their relative performance has remained an open question. 
Here, we clarify this situation.

\subsection{Velocity-Verlet based integration scheme DPD-VV} 
\begin{table}[!]
\hrule
\begin{tabular}{lll}
&   (1) & $\vec{v}_i \longleftarrow \vec{v}_i +
           \frac{1}{2} \frac{1}{m} \left( \vec{F}_i^C \Delta t +
           \vec{F}_i^D \Delta t + \vec{F}_i^R \sqrt{\Delta t} \right)$ \\
&   (2) & $\vec{r}_i \longleftarrow \vec{r}_i + \vec{v}_i \Delta t$ \\
&   (3) & Calculate $\vec{F}_i^C\{\vec{r}_j\}$,
          $\vec{F}_i^D\{\vec{r}_j,\vec{v}_j\}$,
          $\vec{F}_i^R\{\vec{r}_j\}$ \\
&  (4a) & $\vec{v}_i^{\,\circ} \longleftarrow \vec{v}_i +
           \frac{1}{2} \frac{1}{m} \left( \vec{F}_i^C \Delta t +
           \vec{F}_i^R \sqrt{\Delta t} \right)$ \\
&  (4b) & $\vec{v}_i \longleftarrow \vec{v}_i^{\,\circ} +
           \frac{1}{2} \frac{1}{m} \vec{F}_i^D \Delta t$ \\[-4mm]
\setlength{\unitlength}{0.1em}
\begin{picture}(6,30)
 \put(0,3){\line(1,0){6}}
 \put(0,3){\line(0,1){23}}
 \put(0,26){\vector(1,0){6}}
\end{picture}
&   (5) & Calculate $\vec{F}_i^D\{\vec{r}_j,\vec{v}_j\}$ \\
&   (6) & Calculate physical quantities
\end{tabular}
\hrule
\caption{Update scheme for DPD--VV and its self-consistent 
  version SC--VV. In the case of DPD--VV, steps (4b) and (5) 
  are done only once during a single time step. For the 
  self-consistent integrator SC--VV, the loop over steps 
  (4b) and (5) is repeated until the instantaneous 
  temperature has converged to its limiting value. 
  }
\label{table_dpdvv_scvv}
\end{table}

In Table~\ref{table_dpdvv_scvv} we summarize the simplest integrator 
tested in this study. DPD--VV \cite{Bes00,Vat02} is based on the 
standard molecular dynamics 
velocity-Verlet algorithm \cite{Fre02,Ver67,All93} 
which is a time-reversible and symplectic second-order integration 
scheme. These properties can be proven by a straightforward 
application of the Trotter expansion~\cite{Fre02}.
The standard velocity-Verlet
has been shown to be relatively accurate in typical MD 
simulations especially at large time steps \cite{All93}. The 
simplicity and good overall performance of the velocity-Verlet 
algorithm thus makes it a good starting  point for further 
development.

We use the acronym DPD--VV for the modified velocity-Verlet. 
DPD--VV differs from the standard velocity-Verlet scheme in 
one important respect. As discussed above, the dissipative 
forces in DPD depend on the velocities, which in turn are 
governed by the dissipative forces [see Eqs.~(\ref{eq:frij}) 
and (\ref{eq:newton2})]. This matter is not accounted for by 
the standard velocity-Verlet scheme. The DPD--VV, however, accounts 
for this complication in an approximate fashion by updating the 
dissipative forces [step (5) in Table~\ref{table_dpdvv_scvv}] 
for a second time at the end of each integration step. This 
improves its performance considerably yet keeping it computationally 
efficient since the additional update of dissipative forces 
is not particularly time-consuming. In previous studies, the DPD--VV 
scheme has shown good overall performance \cite{Bes00,Vat02} 
for which reason we have chosen it as the ``minimal standard'' 
to which other integrators are compared.

\subsection{Self-consistent velocity-Verlet integrator} 

The update scheme of a self-consistent variant of DPD--VV 
is presented in Table~\ref{table_dpdvv_scvv}. This SC--VV 
algorithm \cite{Bes00,Vat02} determines the velocities 
and dissipative forces self-consistently through 
functional iteration, and the convergence of the iteration 
process is monitored by the instantaneous temperature $k_BT$.
This approach is similar in spirit to the self-consistent 
leap-frog scheme introduced recently by Pagonabarraga {\it et al.} 
\cite{Pag98}, which is the only other published 
self-consistent DPD integration scheme in addition to SC--VV
(to the authors' knowledge). The SC--VV scheme has the 
advantage of being very easy to implement as seen from
Table~\ref{table_dpdvv_scvv}.
A recent study of the SC--VV scheme confirmed that 
it is a promising approach
for interacting DPD systems \cite{Vat02}. 
That is particularly the case for the structural properties
using long time steps in dense systems. 
As a drawback, the SC--VV has no advantage in temperature 
control as compared to other methods. For that, 
there exists
a variant of the SC--VV integrator with a Nos\'e-Hoover type
additional thermostat~\cite{Bes00,Vat02}.

\subsection{Integrator by den Otter and Clarke} 

In 2001 den Otter and Clarke \cite{Ott01} proposed an approach 
which uses a leap-frog algorithm with predefined variables 
$\alpha$ and $\beta$. The idea in the OC-integrator, as it 
is called here, is to try to determine these parameters such 
that the effects due to the velocity dependence of dissipative 
forces are reduced as much as possible. The OC algorithm 
\cite{implementation_of_oc} is described in 
Table~\ref{table_denotter}, in which 
the parameters $\alpha$ and $\beta$ describe the relative 
weight of the random forces with respect to the dissipative 
and conservative ones. They are determined prior to the 
actual DPD simulation by calculating the averages 
$\langle\vec{F}_i^D \cdot \vec{v}_i\rangle$, 
$\langle\vec{F}_i^D \cdot \vec{F}_i^D\rangle$, and 
$\langle\vec{F}_i^R \cdot \vec{F}_i^R\rangle$ from an ensemble 
in which both the kinetic and configurational temperature equal 
the desired temperature $k_B T^{\ast}$ \cite{Ott01}. Once they 
have been calculated, one obtains $\alpha$ and $\beta$ with 
a desired time step $\Delta t$ from the equations
  \begin{equation}
  \label{eq:alp}
  \alpha = \frac{1}{G\Delta t} (1-e^{-G\Delta t})
  \end{equation}
with $G = -\langle\vec{F}_i^D \cdot \vec{v}_i\rangle/
\langle\vec{v}_i \cdot \vec{v}_i\rangle$, and
  \begin{equation}
  \label{eq:bet}
  \beta^2 = - \frac{ 2m\alpha \langle \vec{F}_i^D \cdot \vec{v}_i \rangle
  - \alpha^2 \Delta t \, \langle \vec{F}_i^D \cdot \vec{F}_i^D \rangle} 
                 {\langle \vec{F}_i^R \cdot \vec{F}_i^R \rangle} . 
  \end{equation}
Note that both $\alpha$ and $\beta$ depend explicitly on $\Delta t$. 
Since the averages $\langle\vec{F}_i^D \cdot \vec{v}_i\rangle$, 
$\langle\vec{F}_i^D \cdot \vec{F}_i^D\rangle$, and 
$\langle\vec{F}_i^R \cdot \vec{F}_i^R\rangle$ can be derived 
analytically only for a limited number of systems, one usually 
has to calculate them from simulation (with a very small 
time step). This might be a problem in cases where 
properties such as density and temperature are varied 
over a wide range, since the parameters $\alpha$ and 
$\beta$ should (at least in principle) be calculated separately 
for all different conditions. Nevertheless, den Otter and Clarke 
have shown \cite{Ott01} that the OC algorithm performs well in both 
the ideal gas and a softly interacting DPD fluid. 
\begin{table}[!]
\hrule
\begin{tabular}{ll}
(1) & $\vec{v}_i \longleftarrow \vec{v}_i +
      \alpha \frac{1}{m} \left( \vec{F}_i^C \Delta t +
      \vec{F}_i^D \Delta t \right) +
      \beta \frac{1}{m} \vec{F}_i^R \sqrt{\Delta t}$ \\
(2) & $\vec{r}_i \longleftarrow \vec{r}_i + \vec{v}_i \Delta t$ \\
(3) & Calculate $\vec{F}_i^C\{\vec{r}_j\}$,
      $\vec{F}_i^D\{\vec{r}_j,\vec{v}_j\}$,
      $\vec{F}_i^R\{\vec{r}_j\}$ \\
(4) & Calculate physical quantities
\end{tabular}
\hrule
\caption{The approach OC by den Otter and Clarke. 
    Initialization: Calculate averages 
    $\langle\vec{F}_i^D \cdot \vec{v}_i\rangle$, 
    $\langle\vec{F}_i^D \cdot \vec{F}_i^D\rangle$ 
    and $\langle\vec{F}_i^R \cdot \vec{F}_i^R\rangle$ either 
    analytically or numerically. Then extract $\alpha$ and 
    $\beta$ from Eqs.~(\ref{eq:alp}) and (\ref{eq:bet}), 
    respectively. }
\label{table_denotter}
\end{table}

%
%
\begin{table}[!]
\hrule
\begin{tabular}{l c c c}
\hline\hline 
\multicolumn{1}{c}{Parameter} & 
\multicolumn{1}{c}{Model A}  & 
\multicolumn{1}{c}{Model B}  & 
\multicolumn{1}{c}{Model polymer system} \\
\hline 
$\langle\vec{F}_i^D \cdot \vec{v}_i\rangle  $ & $-7.528088$  & $-4.985245$  &  
$-7.566508$ \\ 
$\langle\vec{F}_i^D \cdot \vec{F}_i^D\rangle$ & 38.254933  & 14.507340  &  
38.532832 \\ 
$\langle\vec{F}_i^R \cdot \vec{F}_i^R\rangle$ & 15.072610  & 9.976453   & 
15.150489  \\
\hline\hline
\end{tabular}
\hrule
\caption{The values of parameters used in the OC integrator.}
\label{oc_parameters}
\end{table}

For the three models considered in this work, we determined 
the parameters $\langle\vec{F}_i^D \cdot \vec{v}_i\rangle$, 
$\langle\vec{F}_i^D \cdot \vec{F}_i^D\rangle$, and 
$\langle\vec{F}_i^R \cdot \vec{F}_i^R\rangle$ separately 
for all cases. Their values are shown in Table~\ref{oc_parameters}.

Note that the parameters for the model polymer system in 
Table~\ref{oc_parameters} have been determined by averaging over 
all particles in the system. An alternate way would be to find 
$(M + 1)$ different values for $\alpha$ and $\beta$ by averaging 
over the solvent and monomer particles separately. However, this 
would require a major computational study prior to actual 
simulations and is therefore not feasible. Besides, it might be 
against the usual spirit as integration schemes are typically 
based on prefactors that are identical for all particles in a system.

\subsection{Shardlow's splitting method}

\begin{table*}[!]
\hrule
  \begin{tabular}{ll}
    (1) & For all pairs of particles for which $r_{ij}<r_c$ \\
        & \begin{tabular}{ll}
           (i)   & $\vec{v}_i \longleftarrow \vec{v}_i -
                    \frac{1}{2} \frac{1}{m} \gamma \omega^2(r_{ij})
                    (\vec{v}_{ij}\cdot\vec{e}_{ij}) \vec{e}_{ij} \Delta t +
                    \frac{1}{2} \frac{1}{m} \sigma \omega(r_{ij}) \xi_{ij}
                    \vec{e}_{ij} \sqrt{\Delta t}$ \\
           (ii)  & $\vec{v}_j \longleftarrow \vec{v}_j +
                    \frac{1}{2} \frac{1}{m} \gamma \omega^2(r_{ij})
                    (\vec{v}_{ij}\cdot\vec{e}_{ij}) \vec{e}_{ij} \Delta t -
                    \frac{1}{2} \frac{1}{m} \sigma \omega(r_{ij}) \xi_{ij}
                    \vec{e}_{ij} \sqrt{\Delta t}$ \\
           (iii) & $\vec{v}_i \longleftarrow \vec{v}_i +
                    \frac{1}{2} \frac{1}{m} \sigma \omega(r_{ij}) \xi_{ij}
                    \vec{e}_{ij} \sqrt{\Delta t} -
                    \frac{1}{2} \frac{1}{m}
                    \frac{\gamma \omega^2(r_{ij}) \Delta t}{1 + \gamma
                    \omega^2(r_{ij}) \Delta t} \left[ (\vec{v}_{ij}\cdot
                    \vec{e}_{ij}) \vec{e}_{ij} + \sigma \omega(r_{ij})
                    \xi_{ij} \vec{e}_{ij} \sqrt{\Delta t} \right]$ \\
           (iv)  & $\vec{v}_j \longleftarrow \vec{v}_j -
                    \frac{1}{2} \frac{1}{m} \sigma \omega(r_{ij}) \xi_{ij}
                    \vec{e}_{ij} \sqrt{\Delta t} +
                    \frac{1}{2} \frac{1}{m}
                    \frac{\gamma \omega^2(r_{ij}) \Delta t}{1 + \gamma
                    \omega^2(r_{ij}) \Delta t} \left[ (\vec{v}_{ij}\cdot
                    \vec{e}_{ij}) \vec{e}_{ij} + \sigma \omega(r_{ij})
                    \xi_{ij} \vec{e}_{ij} \sqrt{\Delta t} \right]$
          \end{tabular} \\
    (2) & $\vec{v}_i \longleftarrow \vec{v}_i +
    \frac{1}{2}\frac{1}{m} \vec{F}_i^C\Delta t$ \\
    (3) & $\vec{r}_i \longleftarrow \vec{r}_i +
    \vec{v}_i\Delta t$ \\
    (4) & Calculate $\vec{F}_i^C\{\vec{r}_j\}$ \\
    (5) & $\vec{v}_i \longleftarrow \vec{v}_i +
    \frac{1}{2}\frac{1}{m} \vec{F}_i^C\Delta t$ \\
    (6) & Calculate physical quantities
  \end{tabular}
\hrule
  \caption{The approach S1 by Shardlow.}
  \label{table_shardlow}
\end{table*}

The most recent addition to DPD integrators has been 
introduced by Shardlow \cite{Sha01}. He applied ideas commonly 
used in solving differential equations to the case of integrating 
the equations of motion in DPD. The key idea is to factorize 
the integration process such that the conservative forces 
are calculated separately from the dissipative and random 
terms. After this splitting the conservative part can be
solved using traditional molecular dynamics methods, while  
the fluctuation-dissipation part is solved separately as
a stochastic differential (Langevin) equation. To this end, 
Shardlow suggested two integrators, called S1 and S2, based 
on splitting the equations of motion up to first and second 
order, respectively.

The formal approach involves a first order splitting 
using the Trotter expansion~\cite{Fre02,tuckerman92} 
(integrator S1) and a second order splitting using the 
Strang expansion \cite{strang} (integrator S2). The 
mathematical details of the derivation can be found in 
Sharlow's original article \cite{Sha01}. It is important to 
notice that the power of the  Trotter (Strang) expansion 
lies in the fact that it provides a general method for 
deriving symplectic algorithms. Importantly, the method 
works for both Hamiltonian and non-Hamiltonian systems; 
see Ref.~\cite{tuckerman92} for a detailed discussion of 
the Trotter expansion and its applications.

Based on our simulation studies and the results presented 
in Ref.~\cite{Sha01} for a system related to the model B in the 
present work, the performance of both of the two splitting methods 
was found to be excellent with S2 displaying slightly better
overall characteristics. Since the first order method (S1) 
is more efficient, we have chosen it to be on the spotlight. 
Here we use the same naming convention and call it S1. The 
algorithm is presented in Table~\ref{table_shardlow}. To the 
best of our knowledge, further tests of S1 have not been 
reported yet.

The curious fact that the algorithm (Table~\ref{table_shardlow})
appears asymmetric for particles $i$ and $j$ is a result of the 
use of the fluctuation-dissipation theorem and Newton's third 
law. The form in which the algorithm is presented keeps the 
notation otherwise symmetric.

\subsection{Lowe's approach --  Lowe-Andersen method}
\label{sect:lowes_integrator}

The approach introduced by Lowe in 1999 \cite{Low99} is presented in 
Table~\ref{table_lowe}, which shows how one first integrates 
the Newton's equations of motion with a time step $\Delta t$, 
and then thermalizes the system as follows. For all pairs of 
particles for which $r_{ij} < r_c$, one decides with 
a probability $\Gamma \Delta t$ whether to take a new relative 
velocity from a Maxwell distribution. For each pair of particles 
whose velocities are to be thermalized, one works on the component 
of the velocity parallel to the line of centers and generates 
a relative velocity $\vec{v}_{ij}^{\,0} \cdot \vec{e}_{ij}$ from 
a distribution $\xi_{ij}^{(g)} \sqrt{2k_BT^{\ast}/m}$. Here 
$\xi_{ij}^{(g)}$ are Gaussian random numbers with zero mean 
and unit variance. This approach has its origin in 
the Andersen thermostat \cite{And80}, hence the name
Lowe-Andersen method.

The key factor in Lowe's method is the parameter $1 / \Gamma$ which 
describes the decay time for relative velocities. Since the condition 
$0 < \Gamma \Delta t \leq 1$ is obvious, one finds that for 
$\Gamma \Delta t = 1$ the particle velocities are thermalized 
at every time step, while for $\Gamma \Delta t \approx 0$ 
the model system is only weakly coupled to the thermostat. 
Thus the dynamical properties of the system can be tuned by 
the choice of $\Gamma$ as shown by Lowe \cite{Low99}.

Although the present version of the algorithm follows the 
original one \cite{Low99} and is based on the velocity-Verlet 
scheme, it is clear that other approaches such as the leap-frog 
are equally useful, if desired. Further, based on the work by 
Lowe \cite{Low99}, this approach seems very promising although 
it has received only little attention by far \cite{Low99b}.

In the present work for the three model systems considered 
here, we set $\Gamma$ such that the tracer diffusion properties 
of the fluid are similar with those of DPD systems with chosen 
$\sigma$ in the limit $\Delta t \rightarrow 0$. In this fashion, 
we end up to a value of $\Gamma = 0.745$ for model A and 
to a value of $\Gamma = 0.44$ for model B. In the model polymer 
system we used the same value as in model A since the solvent 
is described in a similar fashion in both cases. 

\begin{table}[h]
\hrule
  \begin{tabular}{ll}
    (1) & $\vec{v}_i \longleftarrow \vec{v}_i +
    \frac{1}{2}\frac{1}{m} \vec{F}_i^C\Delta t$ \\
    (2) & $\vec{r}_i \longleftarrow \vec{r}_i +
    \vec{v}_i\Delta t$ \\
    (3) & Calculate $\vec{F}_i^C\{\vec{r}_j\}$ \\
    (4) & $\vec{v}_i \longleftarrow \vec{v}_i +
    \frac{1}{2}\frac{1}{m} \vec{F}_i^C\Delta t$ \\
    (5) & For all pairs of particles for which $r_{ij}<r_c$ \\
        & \begin{tabular}{ll}
           (i)   & Generate $\vec{v}_{ij}^{\,\circ}\cdot\vec{e}_{ij}$
                   from a distribution 
                   $\xi_{ij}^{(g)} \sqrt{2k_BT^{\ast}/m}$\\
           (ii)  & $2\vec{\Delta}_{ij} = \vec{e}_{ij}
                   (\vec{v}_{ij}^{\,\circ}-\vec{v}_{ij}) \cdot \vec{e}_{ij}$ \\
           (iii) & $\vec{v}_i \longleftarrow \vec{v}_i +
                   \vec{\Delta}_{ij}$ \\
           (iv)  & $\vec{v}_j \longleftarrow \vec{v}_j -
                   \vec{\Delta}_{ij}$
          \end{tabular} \\
        & with probability $\Gamma\Delta t$ \\
    (6) & Calculate physical quantities
  \end{tabular}
\hrule
  \caption{The approach by Lowe using Gaussian distributed 
      random numbers $\xi_{ij}^{(g)}$ from a 
      distribution $\xi_{ij}^{(g)} \sqrt{2k_BT^{\ast}/m}$.}
  \label{table_lowe}
\end{table}

\section{Performance of integrators}
\label{sec:per}

\subsection{Physical quantities studied}
\label{sec:physical_quantities}

We characterize the integrators by studying a number of physical 
observables. After equilibrating the system, we first calculate 
the average kinetic temperature
  \begin{equation} 
  \langle k_BT \rangle = \frac{m}{3N-3} 
           \left\langle \sum_{i=1}^N \vec{v}_i^{\,2} \right\rangle,
  \end{equation}
whose conservation is one of the main conditions for reliable 
simulations in the canonical ensemble.

Next, in the cases of models A and B, we examine the radial 
distribution function $g(r)$ \cite{Boo80}  which is one of 
the most central observables in studies of structural 
properties of liquids and solids. For the 
ideal gas (model A), the radial distribution function provides 
an excellent test for the integrators since then $g(r) \equiv 1$ 
at the continuum limit. Therefore, any deviation from unity has 
to be interpreted as an artifact due to the integration scheme 
employed.

In model B, in which conservative interactions are present, 
there are no exact theoretical predictions for $g(r)$ that would 
allow a straightforward comparison of  different integration 
schemes. A comparison is possible, though, in terms of physical 
observables such as the compressibility and the coordination 
number that are based on integrating $g(r)$. In the present 
study, we have chosen to consider the coordination number 
defined as 
\begin{equation}
\label{eq:coordination}
N_c = 4 \pi \rho \int_{0}^{\,\,\,\,r_{1}} {\rm dr} \,g(r) \, r^2 , 
\end{equation} 
where $\rho$ is the particle number density of the system and 
$r_1$ is the radial distance at which $g(r)$ has its first 
minimum after the leading (first) peak.

The radial distribution function reflects equilibrium 
(time-independent) properties of the system. To complement 
the comparison of different integrators, we also consider 
the tracer diffusion coefficient 
  \begin{equation} 
  D_T = \lim_{t \to \infty} \frac{1}{6\, t}  
  \langle [ \vec{r}_i(t) - \vec{r}_i(0) ]^2 \rangle , 
  \label{eq:tracer} 
  \end{equation}
which can provide us with information of possible problems 
on the dynamics of the system. Here $\vec{r}_i(t)$ is the 
position of a tagged particle in models A and B, and the 
mean-square displacement is then averaged over all particles 
in a system to get better statistics for $D_T$. In the model 
polymer system, $\vec{r}_i(t)$ describes the center-of-mass 
position of the polymer chain via 
\begin{equation} 
\vec{r}_{\rm cm}(t) = 
            \frac{ 1 }{ M } \sum_{i=1}^{M} \vec{r}_i(t) \, , 
\end{equation} 
where the index runs over monomers in a chain.

For the model polymer chain, we further calculate the radius 
of gyration $R_g \equiv \sqrt{ \langle R_{g}^2 \rangle }$ 
defined as 
\begin{equation} 
 \label{eq:rg2}
 \langle R_{g}^2 \rangle \equiv
   \frac{1}{M} \sum_{i=1}^{M} \left\langle
   [ \vec{r}_i - \vec{r}_{\rm cm} ]^2 \right\rangle ,
\end{equation} 
which shows that $R_g$ is a measure of polymer size. It is  
actually one of the most central quantities in polymer science 
and therefore serves as an excellent measure for our purposes. 

In the following, the errors are stated in the figure captions
and given as the magnitude of standard deviation.

\subsection{Results for model A} 

\begin{figure}[!]
\centering\epsfig{figure=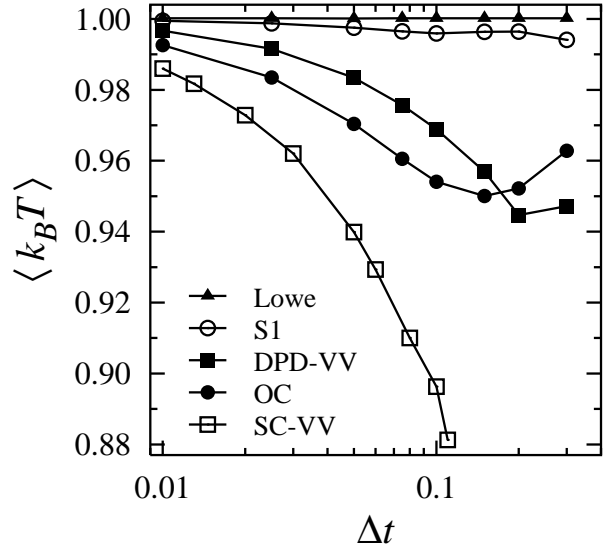,width=11cm}
\caption{
Results for the deviations of the observed temperature 
$\langle k_BT \rangle$ from the desired temperature 
$k_B T^{\ast} \equiv 1$ vs. the size of the time step 
$\Delta t$ in model A.
The error in $\langle k_BT \rangle$ is of the order of $10^{-4}$.} 
\label{figure:model_a_temp}
\end{figure}

\begin{figure*}[!]
\centering\epsfig{figure=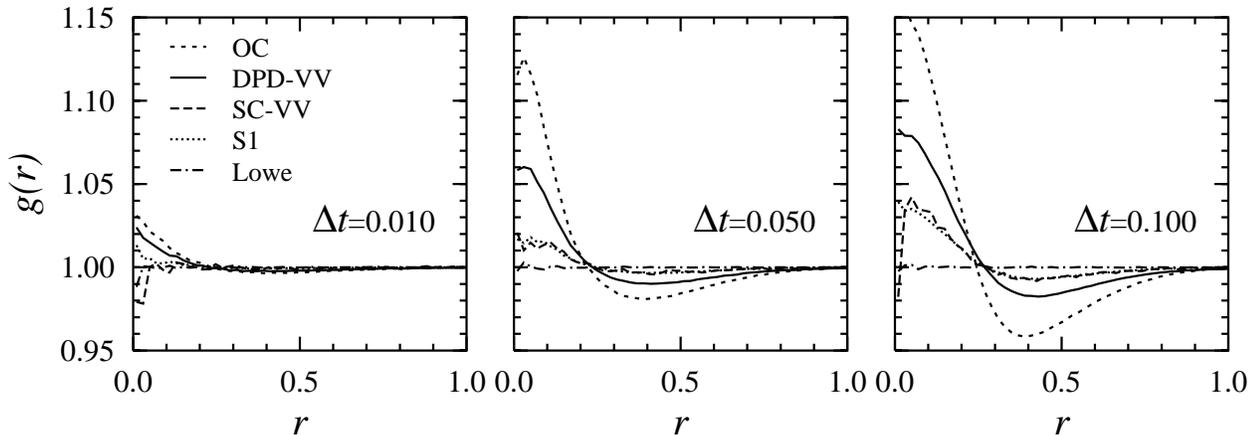,width=17cm}
\caption{
Radial distribution functions $g(r)$ with several values 
of time step $\Delta t$ in model A: (a) $\Delta t = 0.01$, 
(b) $\Delta t = 0.05$, and (c) $\Delta t = 0.1$. 
The error in $g(r)$ is greatest at $r=0.01$, where it takes the 
value of 0.01.
}
\label{figure:model_a_gr}
\end{figure*}

As discussed above, model A is characterized by the absence 
of conservative forces, and thus any artifacts arising from 
the velocity-dependent forces are expected to be pronounced 
in this model. To study this possibility, we first discuss 
the deviations of the observed kinetic temperature 
$\langle k_BT \rangle$ from the desired temperature 
$k_BT^{\ast}$. The results for $\langle k_BT \rangle$ shown 
in  Fig.~\ref{figure:model_a_temp} indicate that DPD--VV and 
SC--VV are reasonably good at small time increments, but larger 
time steps lead to major deviations from the desired temperature. 
For OC, $\langle k_BT \rangle$ decreases monotonically with 
$\Delta t$ for $\Delta t \leq 0.15$, after which the temperature 
increases rapidly. Nevertheless, the deviation is greater 
than in the case of DPD--VV but considerably smaller than 
for SC--VV. The Shardlow S1 integrator, however, has 
very good temperature control and the deviations remain 
less than 0.5\,\% up to $\Delta t = 0.2$. The best temperature 
control is found for the method by Lowe, however, yielding 
$\langle k_BT\rangle = k_BT^{\ast}$ for all time steps 
$\Delta t$. In this case, we have extended the studies 
further and tested the behavior of $\langle k_BT \rangle$ 
with various values of $\Gamma$ between 0.1 and 10, but the 
conclusions remain the same.

Results for $g(r)$ are shown in Fig.~\ref{figure:model_a_gr}. 
We find that the deviations from the ideal gas limit $g(r) = 1$ 
are pronounced for OC, indicating that even for small time 
steps this integration scheme gives rise to unphysical correlations. 
The performance of DPD--VV is considerably better, although 
artificial structures are yet rather pronounced, while SC--VV 
and S1 lead to a radial distribution function that is close to 
the theoretically predicted one. Completely structureless $g(r)$ 
is found only for the integrator by Lowe, however. Again, in 
this case, we have tested the behavior of $g(r)$ with various 
values of $\Gamma$, but the results remain the same. This 
confirms the expectation that $\Gamma$ does not influence 
the equilibrium properties of the system.

\begin{figure}[!]
\centering\epsfig{figure=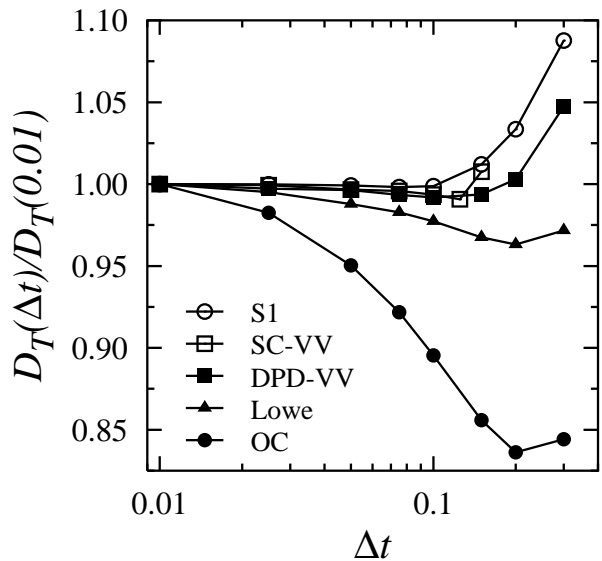,width=11cm}
\caption{
Results for the tracer diffusion coefficient $D_T(\Delta t)/D_T(0.01)$
vs. the time step $\Delta t$ in model A. The error in
$D_T(\Delta t)/D_T(0.01)$ \,is of the order of 0.001.}
\label{figure:model_a_dt}
\end{figure}

The results for the diffusion coefficient $D_T$ in 
Fig.~\ref{figure:model_a_dt} are essentially consistent with 
the conclusions above. The integrator OC is not very useful 
in a system of the present kind, since it seems to lead to 
substantial deviations from the expected behavior. The SC--VV 
and the integrator by Lowe perform much better, while the 
integrators S1 and DPD--VV are most stable in this case.

\subsection{Results for model B} 

\begin{figure}[!]
\centering\epsfig{figure=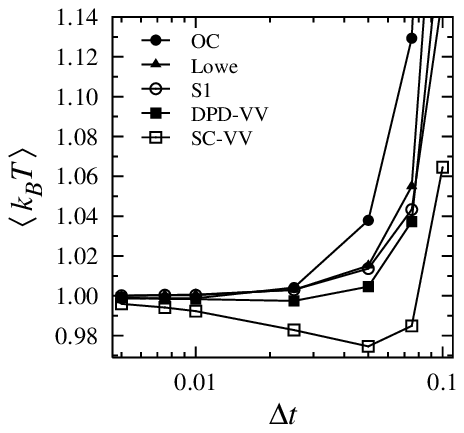,width=11cm}
\caption{
Results for the deviations of the observed temperature 
$\langle k_BT \rangle$ from the desired temperature 
$k_BT^\ast\equiv 1$ vs. the size of the time step 
$\Delta t$ in model B.
The error in $\langle k_BT \rangle$ is of the order of $10^{-4}$.
}
\label{figure:model_b_temp}
\end{figure}

\begin{figure*}
\centering\epsfig{figure=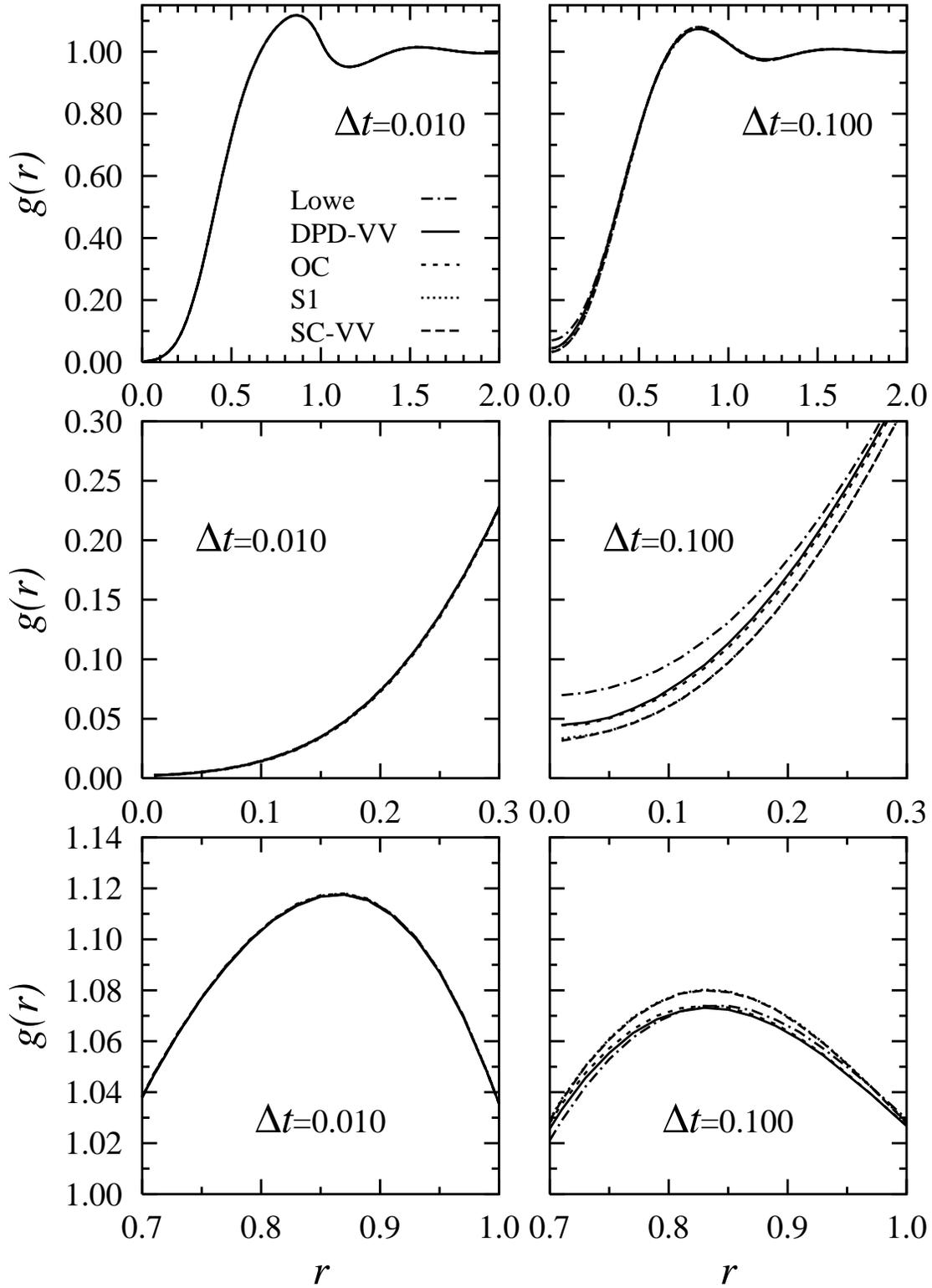,width=15cm}
\caption{
Radial distribution functions $g(r)$ with time 
steps $\Delta t = 0.01$ (on the left) and 
$\Delta t = 0.1$ (on the right) in model B. 
In addition to the full curves, two sets of 
the same data on an expanded scale are also given 
to clarify the deviations between different 
integration schemes.
The error in $g(r)$ is greatest at $r=0.01$, where it takes the
value of 0.001.
}
\label{figure:model_b_gr}
\end{figure*}

Results shown in Fig.~\ref{figure:model_b_temp} for the observed 
kinetic temperature $\langle k_BT \rangle$ reveal that the 
differences between the integration schemes deviations are 
weaker in model B than in model A. As it turns out below, 
this conclusion is generic and holds for all quantities 
studied here.

The deviations of $\langle k_BT \rangle$ from the desired 
temperature $k_BT^{\ast}$ are very minor for all integrators 
at small time steps $\Delta t \leq 0.01$. Differences between 
the integrators become evident only at larger time steps. 
We first find how the temperature in SC--VV first decreases, 
then has a pronounced minimum around $\Delta t \approx 0.05$, 
after which $\langle k_BT \rangle$ increases very rapidly 
and the integrator eventually becomes unstable. The 
temperature conservation of OC is also relatively poor at 
time steps beyond $\Delta t = 0.025$. Best performance in 
this respect is found for the remaining integration schemes 
DPD--VV, S1, and Lowe, whose behavior is quite similar and 
comparable to each other.

Demonstrative results for the radial distribution functions in 
model B are shown in Fig.~\ref{figure:model_b_gr}. It is clear 
that large time steps lead to major problems with regard to pair 
correlations. This is particularly clear at small distances 
($ r < 0.2 $). To quantify these changes, we calculated the 
coordination number [see Eq.(\ref{eq:coordination})]. The results 
shown in Table~\ref{table_coordination} highlight the fact that 
all integration schemes converge to the same result at small 
time steps $\Delta t \leq 0.05$, while for larger time steps 
there are significant deviations from the correct behavior 
found in the limit $\Delta t \rightarrow 0$. However, it is 
somewhat surprising that the coordination numbers obtained 
by different integration schemes are essentially similar 
within error limits, while based on $g(r)$'s there are 
noticeable differences between the pair correlation 
properties of different integrators. This is a consequence of the 
fact that due to the integration in Eq.~(\ref{eq:coordination}), 
the deviations in $g(r)$ toward too small and too large values 
compensate each other. A similar effect has been observed 
recently in compressibility \cite{Vat02}, which is also 
defined as an integral over $g(r)$.

\begin{figure}[!]
\centering\epsfig{figure=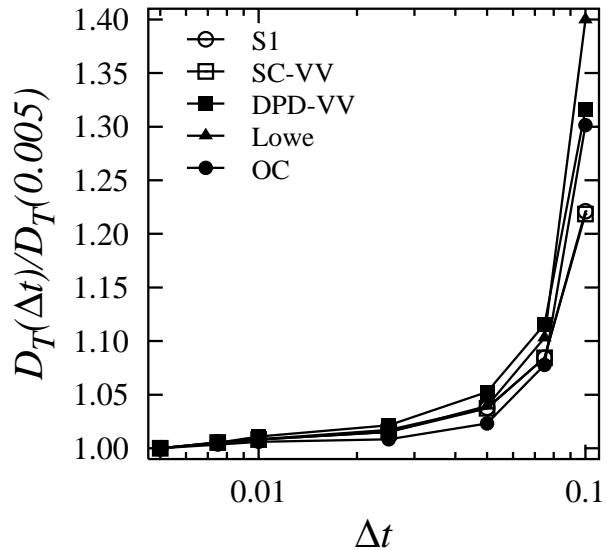,width=11cm}
\caption{
Results for the tracer diffusion coefficient $D_T(\Delta t)/D_T(0.005)$
vs. the time step $\Delta t$ in model B. The error in
$D_T(\Delta t)/D_T(0.005)$ \,is of the order of 0.001.}
\label{figure:model_b_dt}
\end{figure}

Tracer diffusion data shown in Fig.~\ref{figure:model_b_dt} is 
consistent with the conclusions above. For small time steps the 
results of all integration schemes are comparable, while for 
large time steps we can find how the differences become more 
and more pronounced. The scatter in the data does not allow us 
to make conclusive statements of the relative merits of the 
integrators, however. 
\begin{table*}[!]
\begin{tabular}{lccccc}
\hline\hline 
\multicolumn{1}{c}{Integrator} & 
\multicolumn{1}{c}{$\Delta t = 0.005$}  & 
\multicolumn{1}{c}{$\Delta t = 0.010$}  & 
\multicolumn{1}{c}{$\Delta t = 0.050$}  & 
\multicolumn{1}{c}{$\Delta t = 0.075$}  & 
\multicolumn{1}{c}{$\Delta t = 0.100$}  \\
\hline 
DPD--VV & 25.36 & 25.37 & 25.74 & 26.72 & 28.98 \\
SC--VV  & 25.43 & 25.54 & 25.62 & 26.78 & 28.57 \\
OC      & 25.46 & 25.51 & 25.71 & 26.69 & 28.74 \\
S1      & 25.53 & 25.41 & 25.73 & 26.71 & 28.44 \\
Lowe    & 25.35 & 25.42 & 25.59 & 26.95 & 29.12 \\
\hline\hline
\end{tabular}
\caption{Results for the coordination number $N_c$ in model B.
         Error bars are about $\pm 0.05$.}
\label{table_coordination} 
\end{table*}

\subsection{Results for the model polymer system}

Before we consider the results for the model polymer 
system, we would like to emphasize certain similarities 
it has with model A. Namely, in both cases the solvent 
is an ideal gas governed by dissipative and random 
forces only. Further, the model polymer system is dilute 
(more than 99.5\% of the particles in a system are solvent 
particles) and there are no conservative interactions between 
the monomers and the solvent particles. This suggests that 
the artifacts due to integration schemes in the model polymer 
system would be essentially similar to those found in model A. 
It turns out below, however, that this is not the case. 

\begin{figure*}[!]
\centering\epsfig{figure=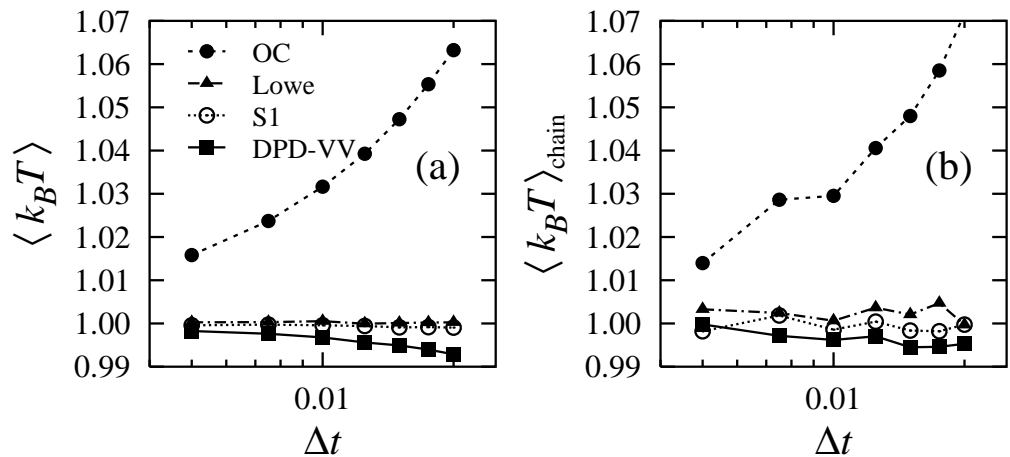,width=14cm}
\caption{
Results for the deviations of the observed temperature 
$\langle k_B T \rangle$ from the desired temperature 
$k_BT^\ast\equiv 1$ vs. the size of the time step 
$\Delta t$ in the model polymer system. In (a) we show 
the observed temperature of the whole system, while the 
results in (b) correspond to the polymer chain only 
(see text for details).
For the whole system, the error is of the order of $10^{-4}$,
while for the polymer chain the error is 0.004.
}
\label{figure:temp_polymer} 
\end{figure*}

Results shown in Fig.~\ref{figure:temp_polymer}(a) indicate 
that the observed kinetic temperature $\langle k_BT \rangle$ 
only rather weakly depends on the integration scheme. Results 
for S1, Lowe, and DPD--VV are all within one percent up to 
$\Delta t \approx 0.02$ above which the integration schemes 
become unstable. The OC integrator is somewhat less reliable 
in this case, as it leads to a monotonous increase of 
$\langle k_B T \rangle$, thus differing rather clearly from 
the results of other integration schemes.

A comparison of Figs.~\ref{figure:model_a_temp} and 
\ref{figure:temp_polymer}(a) provides one with an intriguing 
view of effects that arise from the hybrid approach. We 
first note that Lowe's approach as well as S1 are 
equally good, in agreement with the conclusions made 
in model A. However, in model A the integration schemes were 
found to be stable up to very large time steps on the order 
of $\Delta t \approx 0.4$, while in the model polymer system 
the largest time steps possible are about 0.02. This is due 
to the hard conservative monomer-monomer interactions used in 
describing the polymer chain, for which reason the size of the 
time step has to be reduced considerably as compared to model A. 
This suggests that, for practical purposes in hybrid models, 
one should seriously consider integration schemes with two 
different time steps, one for the solvent and another for the 
polymer degrees of freedom. Another interesting feature 
concerns the behavior of $\langle k_BT \rangle$ in the case of 
OC. In model A, the observed kinetic temperature decreased 
monotonously down to timesteps $\Delta t \approx 0.15$, while in the 
model polymer system the trend is the opposite. This is consistent 
with our results for model B, and suggests that the artifacts 
due to integration schemes depend significantly on the 
interactions chosen for the system.

To gain further insight into the performance of the integration 
schemes in a hybrid approach, we studied a number of physical 
quantities that specifically characterize the properties of 
the polymer chain. First, we studied the observed kinetic 
temperature of the {\it polymer}, defined as 
  \begin{equation} 
  \langle k_BT \rangle_{\rm chain} = \frac{m}{3M} 
           \left\langle \sum_{i=1}^M \vec{v}_i^{\,2} \right\rangle . 
  \end{equation}
This quantity characterizes the thermal fluctuations of the 
polymer chain, and therefore if there are any serious problems 
due to the integration schemes, then we expect that they are 
manifested in the behavior of $\langle k_BT \rangle_{\rm chain}$.

We find that the results in Fig.~\ref{figure:temp_polymer}(b) 
are wholly consistent with those presented in 
Fig.~\ref{figure:temp_polymer}(a). Essentially, this implies 
that the temperature deviations in {\it the whole system} 
arise from the hard interparticle interactions used to describe 
{\it the solute} even though the system is dilute. We think 
that this finding is of generic nature and applies to both 
hybrid models and other DPD simulations in which all 
interactions are approximately of equal magnitude. In 
particular, it allows us to suggest that the artifacts due 
to integration schemes in DPD are predominated by {\it the 
interactions that dictate the size of the time step}.

In systems with hard conservative interactions 
the time step must be small. Otherwise, gradients in forces 
become too large and the system becomes unstable. Consequently, 
if the conservative force is stronger than the dissipative 
and random forces  the artifacts due to integration 
schemes are driven by the conservative forces just like in
classical molecular dynamics simulations. Naturally, the
velocity-dependent dissipative forces are still playing 
a role but their effect is not as important as the influence 
of conservative interactions. From a practical point of view, 
this means that there is no particular reason to use an 
integrator which accounts for the velocity dependence of 
dissipative forces.

On the other hand, if all interactions are soft, or if the 
conservative forces are weak compared to the dissipative forces, 
then it is plausible that the velocity-dependence of dissipative 
forces is the underlying reason for artifacts due to the 
integration procedure. This is the case when the time step 
is determined by the dissipative force rather than the 
hard-core of the conservative potential. In this situation 
the quality of the integration scheme is very important, and 
the velocity dependence of the dissipative forces has to be 
accounted for by the integration scheme. It is clear that 
this matter warrants attention and should be accounted for 
in all subsequent studies by DPD.

Further studies of the radius of gyration and the diffusion 
coefficient of the polymer chain revealed the expected and 
undesired fact that computational studies of a dilute polymer 
system in an explicit solvent are very time consuming, and 
therefore the error bars remained rather large despite major 
computational efforts. Consequently, we found that the results 
for $R_g$ and $D_T$ of different integrators were essentially 
equal within error limits (data not shown). It is likely that 
more extensive calculations would have expressed deviations 
between different integration schemes, but we concluded that 
such studies were not worthwhile.

\subsection{Computational efficiency}

Besides the strength of the artifacts, we have paid 
attention to the computational efficiency of the integration 
schemes by calculating the cpu time needed for a single time 
step $\Delta t$. Although this depends on various 
practical matters such as the computer architecture, the 
implementation of the algorithms, and the size of Verlet 
neighbor tables, we think this approach can provide one with 
the essential information of the relative speed of the 
integrators tested in this work.

The tests for efficiency have been carried out on a Compaq 
Alpha Station XP1000 with a 667 MHz processor. To this end,
we used model B (with 4\,000 particles at a density of
$\rho = 4$) with standard Verlet neighbor tables 
\cite{All93} and a time step of $\Delta t = 0.05$.

We calculated the cpu time needed for the integration of 
a single time step (averaged over 1\,000 consecutive steps). 
The times needed to update the Verlet neighbor tables (or 
to calculate any physical quantities) are not included in 
these results. Thus the DPD--VV and SC--VV schemes were 
considered over steps (1)\,--\,(5) in  
Table~\ref{table_dpdvv_scvv}, the OC approach over  
steps (1)\,--\,(3) in Table~\ref{table_denotter},  
Shardlow's approach over steps (1)\,--\,(5) in  
Table~\ref{table_shardlow}, and Lowe's integration 
scheme over steps (1)\,--\,(5) in Table~\ref{table_lowe}.
In the case of SC--VV, steps (4b) and (5) were repeated 
six times which guaranteed self-consistency in this case. 
Note that Lowe's approach is based on normally distributed 
random numbers while other integration schemes use uniformly 
distributed ones. Since there are various methods available 
for the generation of normally distributed random numbers, 
this may affect the efficiency of Lowe's approach to 
some extent \cite{randomnumbers}.

\begin{table}
\hrule
\begin{tabular}{l c}
\hline\hline 
\multicolumn{1}{c}{Integrator} & 
\multicolumn{1}{c}{Cpu time (seconds)} \\
\hline 
DPD--VV & $0.0363 \pm 0.0005$ \\ 
SC--VV  & $0.1010 \pm 0.0009$ \\ 
OC      & $0.0251 \pm 0.0005$\\  
S1      & $0.0256 \pm 0.0005$\\  
Lowe    & $0.0143 \pm 0.0005$\\ 
\hline 
Verlet list  & $0.0293 \pm 0.0003$\\
\hline\hline
\end{tabular}
\hrule
\caption{Results for the computational efficiency of the 
integration schemes. Shown here are results for integrating 
the equations of motion over one time step of size $\Delta t = 0.05$, 
although the results have been averaged over 1\,000 consecutive 
steps. For the purpose of comparison, the time needed to update 
the Verlet neighbor list has also been given; the time shown 
here corresponds to its minimum value when the list is small 
since it is updated after every time step. 
(Simulation parameters: $k_B T = 1$, ${\mathcal A} = 25$, $\sigma = 3$,
and $\rho = 4$.)} 
\label{table:efficiency} 

\end{table}
The results shown in Table~\ref{table:efficiency} indicate 
that Lowe's method is substantially faster than OC and S1,
which in turn are clearly faster than DPD--VV. Finally, the 
result that SC--VV is considerably slower than DPD--VV is 
not surprising due to the iteration process for the velocities 
and dissipative forces.

When these times are compared to each other, one should also 
keep in mind that DPD--VV, S1, and Lowe's method are 
significantly easier to deal with compared to SC--VV and OC.
In SC--VV, one needs to find the sufficient number of iterations 
prior to actual simulations to guarantee self-consistency.
In OC, the preliminary work required prior to simulations
is even more extensive, as one has to determine the parameters
$\alpha$ and $\beta$ for a given system under desired 
thermodynamic conditions. This task may indeed take some time.

\section{Discussion and Conclusions}
\label{sec:sum}

\begin{figure*}[!]
\centering\epsfig{figure=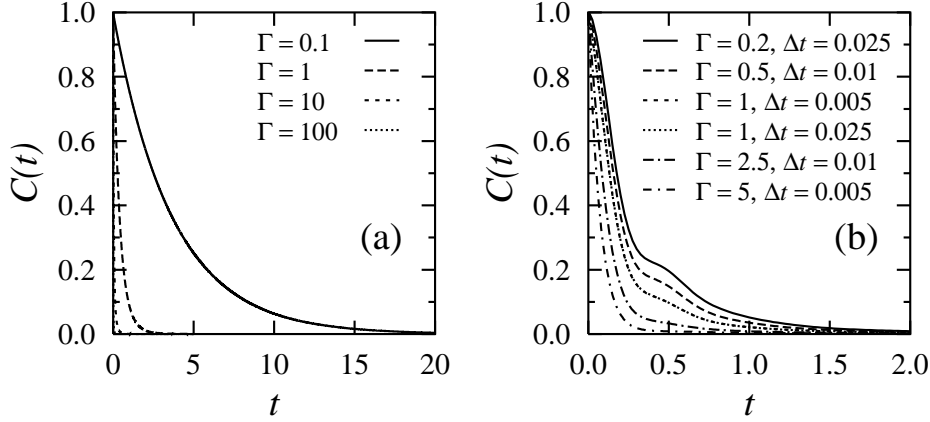,width=15cm}
\caption{
a) Velocity autocorrelation function for Model A in Lowe's method.
The error is of the order of $10^{-4}$.
b) Velocity autocorrelation function for Model B in Lowe's method.
The error is of the order of $10^{-4}$.}
\label{figure:auto1}
\end{figure*}
\begin{figure*}[!]
\centering\epsfig{figure=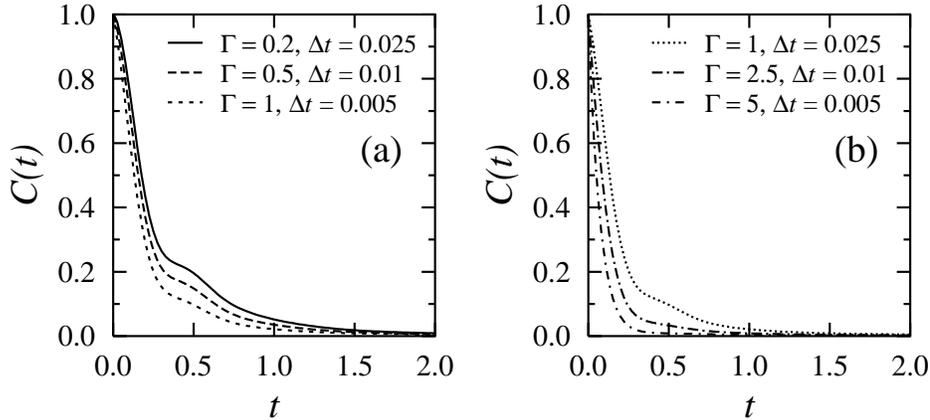,width=15cm}
\caption{
Velocity autocorrelation function for Model B in Lowe's method
with the product $\Gamma \Delta t$ fixed to (a) 0.005 and (b) 0.025. 
The error is of the order of $10^{-4}$.}
\label{figure:auto2}
\end{figure*}

In this work, we have tested several novel schemes on an equal 
footing through DPD simulations of three different model systems. 
The first of the models corresponds to a case where conservative 
interactions play no role, while the second model describes
fluid-like systems with relatively strong but 
soft conservative potentials. Finally, the third model 
aims to characterize the quality of integration 
schemes in a hybrid approach for a dilute polymer system.

Of the integration schemes considered here, 
DPD--VV and SC--VV have recently been examined
in Refs.~\cite{Bes00,Vat02}. The results of the present 
study are consistent with previous findings: DPD--VV
exhibits good overall performance, indicating 
that it presents a relatively accurate means to integrate 
the equations of motion at a reasonable computational 
cost.

Of the previously untested methods
the Otter-Clarke (OC) method \cite{Ott01} is fast, performing 
especially well in interacting systems in which conservative 
forces are important. A drawback is that the 
parameters $\alpha$ and $\beta$ need to be determined prior to 
actual simulations through time-consuming precursory simulations 
with a very small time step. We note, however, that 
the properties of the OC scheme are in fact relatively 
insensitive to slight changes in $\alpha$ and $\beta$. 
Thus, for example, studies of the model polymer system
using specifically determined $\alpha$ and $\beta$ values 
yielded results almost identical with studies based on 
the parameters of model A.

The Shardlow S1 integrator \cite{Sha01} is possibly the 
brightest star in this work. It performed very well in all 
models, and it is fast and rather easy to implement. 
We feel that it presents the best choice of integration  
schemes within  the ``usual'' conceptual framework of DPD.

Interestingly, however, we have also found that the elegant 
and conceptually distinct method of Lowe \cite{Low99}
performed excellently and is easy to implement.
Furthermore, and what is important when 
Lowe's method is compared to Shardlow's integration scheme, 
it provides an {\it alternative} and a very 
attractive description of dissipative particle dynamics. 
Thus, a direct comparison of S1 and Lowe's method is not 
meaningful. Instead, we discuss the pros
and cons of these two approaches.

The usual DPD description is based on the idea that 
soft matter systems can be
described in terms of softly interacting particles
with some of the degrees of freedom 
coarse grained out and replaced with random noise 
coupled to dissipation. 
Temperature conservation is achieved through 
the fluctuation-dissipation theorem
and the correct hydrodynamic behavior is guaranteed
by momentum conservation~\cite{Esp95}. 
Various studies have extended these ideas
further. For example, Flekk{\o}y et al. 
developed a DPD framework starting from a microscopic 
description \cite{Fle99,Fle00}. Espa{\~n}ol and coworkers, 
in turn, studied the dependence of transport properties of 
DPD fluids on the length and time scales \cite{Rip01} and 
a generalization of DPD to energy conserving systems \cite{Esp97}. 
DPD has recently been used together with molecular
dynamics to coarse grain aqueous salt solutions~\cite{Lyu02}  
in which the effective interactions 
used in DPD simulations were obtained from MD simulations 
by the inverse Monte Carlo procedure \cite{Lyu95}. 
The last ten years have been very successful 
on both the analytical and the computational 
fronts--the theoretical basis of DPD is now 
well established, and the number of 
applications has increased at a steady pace.

Lowe's \cite{Low99} approach is a very recent inception and 
thus far has received limited attention. Although the 
theoretical foundations of Lowe's method have yet to be
fully worked out, it offers promising aspects that are not 
obvious in the traditional DPD description. To clarify 
these aspects, let us first remind ourselves that 
Lowe's method does not include dissipation in the 
usual sense. Rather, it is based on a thermostat that 
thermalizes the velocities of pairs of particles at 
a rate which depends on the dynamical parameter $\Gamma$. 
This parameter tunes the dynamical properties of the 
system. Lowe pointed out that the soft interactions used in 
DPD lead to a situation where the ratio of the kinematic 
viscosity and the diffusion coefficient of solvent 
particles (known as the Schmidt number {\it Sc}) is of 
the order of one. This value corresponds to a situation 
often found in gases, while in fluids ${\rm {\it Sc}} \sim 10^3$ 
or even larger. To get closer to more realistic values 
for {\it Sc}, one can reduce the diffusion rate by 
using harder interparticle interactions, but this is
against the philosophy of DPD and would reduce some 
of the benefits of the DPD approach.

Lowe's approach is very different in this respect. 
It allows one to adjust the viscosity of the system to 
a desired value by varying the dynamical parameter $\Gamma$ 
while the diffusive properties are not considerably affected 
since the conservative interparticle interactions remain soft. 
As a result, the Schmidt number can obtain values as large 
as $10^7$ \cite{Low99}. When compared to the 
usual DPD description, this implies that Lowe's approach may 
be more feasible for describing hydrodynamic systems in which 
one needs to worry about the time scales of momentum diffusion 
and mass transfer with respect to the size of the colloidal 
particle.

There still remains the issue of the practical viability
of Lowe's approach, since we are not aware of any 
applications where the method by Lowe has been used. 
However, we are positive that this approach is a promising 
technique. For example, we have recently applied  
Lowe's method to microphase separation of block copolymers 
in the spirit of Groot and Madden \cite{Gro98}, and it 
turned out that Lowe's method was able to reproduce their results.
Finally, in Figs.~\ref{figure:auto1} and \ref{figure:auto2} 
we show how the velocity autocorrelation function depends on 
the choice of $\Gamma$ for models A and B (Fig.~\ref{figure:auto1}), 
and how it is affected when $\Gamma$ is varied but keeping the 
product $\Gamma \Delta t$ constant in the case of model B 
(Fig.~\ref{figure:auto2}). As discussed above, it is clear 
that large values of $\Gamma$ lead to faster decay. However, 
the qualitative behavior of the velocity autocorrelation 
function is not seemingly affected, as illustrated by 
Fig.~\ref{figure:auto2}, and the effect of $\Gamma$ on the 
diffusion coefficient was found not to be important for 
the studied combinations of $\Gamma$ and $\Delta t$.

To conclude, we have studied the performance of various 
novel integration schemes that have been designed 
specifically for DPD simulations. We have tested these 
integration schemes in three different model systems by 
varying the nature of interactions and found that the 
artifacts due to the integration scheme are essentially 
driven by the interactions that dictate the size of the 
time step. Thus, the artifacts and the performance of 
integrators are model dependent. Overall, we have found 
that there are two approaches whose performance is above 
the others. Of these, Shardlow's integration scheme 
is based on splitting the equations of motion and can be 
applied to the usual DPD picture, while the approach by 
Lowe is distinctly different in nature and is related to 
the classical work by Andersen.

\bigskip

\bigskip

\acknowledgements

We would like to thank Gerhard Besold, Wouter den Otter, Alex Bunker, 
and James Polson for useful discussions at the early stages 
of this work, Tony Shardlow for sharing his results prior 
to publication, and Nick Braun for a critical reading of the 
manuscript. This work has, in part, been supported 
by the Academy of Finland (I.V.) and by the Academy of 
Finland Grant No.~54113 (M.K.).


\begin{thebibliography}{250}
\bibitem{implementation_of_oc} 
We have also examined a slightly different implementation 
of the OC scheme in which the velocities are updated in two 
parts as in DPD--VV. This choice may sound more appropriate 
as it comes naturally from the Trotter expansion 
\protect\cite{Fre02}. However, we found that it has only 
a minor effect on the results and was therefore not 
discussed in more detail in this work.
\bibitem{randomnumbers} 
In the present work, we simplified the analysis of the 
efficiency by generating normally distributed numbers 
by the central limit theorem. If needed, this approach 
allows one to decompose the contribution due to random 
number generation, since it involves only a minimal 
number of machine-specific functions. Then, $i=1,2,\ldots,n$ 
uniformly distributed random numbers $x_i \in [0,1]$ 
were summed together to find a single normally distributed 
random number $y = (\sum_{i=1}^{n} x_i - n/2) / \sqrt{12/n}$. 
Here we used $n = 12$ in which case deviations from the 
actual normal distribution are relatively weak.
\end{thebibliography}
\end{document}